\documentclass{article}
\PassOptionsToPackage{numbers,compress}{natbib}
\usepackage[preprint]{neurips_2025}
\usepackage[utf8]{inputenc}
\usepackage[T1]{fontenc}
\usepackage{float}  %
\usepackage{hyperref}
\usepackage{url}
\usepackage{booktabs}
\usepackage{amsmath}
\usepackage{amsfonts}
\usepackage{nicefrac}
\usepackage{microtype}
\usepackage{xcolor}
\usepackage{graphicx}
\usepackage{adjustbox}
\usepackage{etoolbox}
\AtBeginEnvironment{table}{\small\setlength{\tabcolsep}{4pt}}

\newcommand{\siteurl}{https://monperrus.github.io/boot-to-claude-website}
\newcommand{\nprompts}{28}
\newcommand{\ngoaheads}{14}
\newcommand{\ndismissed}{1}
\newcommand{\napprovals}{2}
\newcommand{\plantoosmin}{35}
\newcommand{\nquestions}{8}
\newcommand{\nrecommended}{7}
\newcommand{\ncompactions}{3}
\newcommand{\nassistant}{4\,161}
\newcommand{\ntoolcalls}{2\,214}
\newcommand{\nbash}{1\,272}
\newcommand{\nedit}{551}

\newcommand{\tcactivehours}{6.2}

\newcommand{\ncommits}{23}
\newcommand{\sessionmodel}{\texttt{claude-sonnet-5}}
\newcommand{\ospin}{\texttt{a4a64cb}}
\newcommand{\gamerules}{409}
\newcommand{\gamegui}{212}
\newcommand{\gamesprites}{81}
\newcommand{\vgadriver}{327}
\newcommand{\fontlines}{111}
\newcommand{\inputlines}{21}
\newcommand{\noraclesruns}{125}
\newcommand{\noraclesfailed}{23}
\newcommand{\nostestruns}{51}

\newcommand{\ntoolchaintestruns}{64}
\newcommand{\ntoolchainfixtures}{16}

\newcommand{\ntdefects}{127}
\newcommand{\ntproduct}{113}
\newcommand{\ntoracle}{12}
\newcommand{\ntenv}{2}
\newcommand{\ntreported}{59}
\newcommand{\ntunreported}{54}
\newcommand{\ntcc}{57}

\newcommand{\nttoolchain}{91}
\newcommand{\ntbuildrevealed}{40}
\newcommand{\ntbootrevealed}{18}
\newcommand{\ntcodereading}{34}
\newcommand{\ntsilent}{45}
\newcommand{\ntmaxattempts}{14}
\newcommand{\ntmultiattempt}{31}
\newcommand{\ntsevenfailed}{14}
\newcommand{\ntfirstinrun}{25}
\newcommand{\ntconfirmedinrun}{32}
\newcommand{\playruns}{16}
\newcommand{\playseconds}{307}

\newcommand{\sysname}{TinyMonOS}
\hypersetup{pdftitle={\sysname: An Operating System Entirely Written by AI Built by a Compiler Entirely Written by AI},
  pdfauthor={Martin Monperrus}}
\newcommand{\defect}[2]{\href{\siteurl/defects.html\##1}{#2}}

\title{\sysname{}: An Operating System\\ Entirely Written by AI\\ Built by a Compiler\\ Entirely Written by AI}
\author{%
  Martin Monperrus\\
  KTH Royal Institute of Technology\\
  \texttt{monperrus@kth.se}
}

\begin{document}
\maketitle

\begin{abstract}
We report on \sysname{}, an x86 operating system in which every line of code, from the
boot sector to the compiler that builds it, was written by Claude working as a coding
agent. \sysname{} boots in QEMU through its own two-stage boot loader into a 32-bit
kernel with a disk driver, a file system, a shell, a scripting language, and a compiler. Its build uses no third-party compiler, assembler or linker:
the agent wrote its own, and bootstrapped them until the second and third self-compiled
generations were byte-identical. The human operator set the goals and wrote no code. 
What made this possible is the oracles. The agent combined expected-output, differential
and fixed-point oracles and ran them \noraclesruns{} times; \ntsilent{} of the \ntproduct{}
defects in the system raised no error at all, and only a test booting the whole machine caught
the subtlest ones. We distill guidelines for designing oracles for coding agents. The code
is at \url{https://github.com/monperrus/TinyMonOS} and the full transcripts are
browsable at \url{\siteurl}.
\end{abstract}

\begin{figure}[H]
\centering
\includegraphics[width=0.75\textwidth]{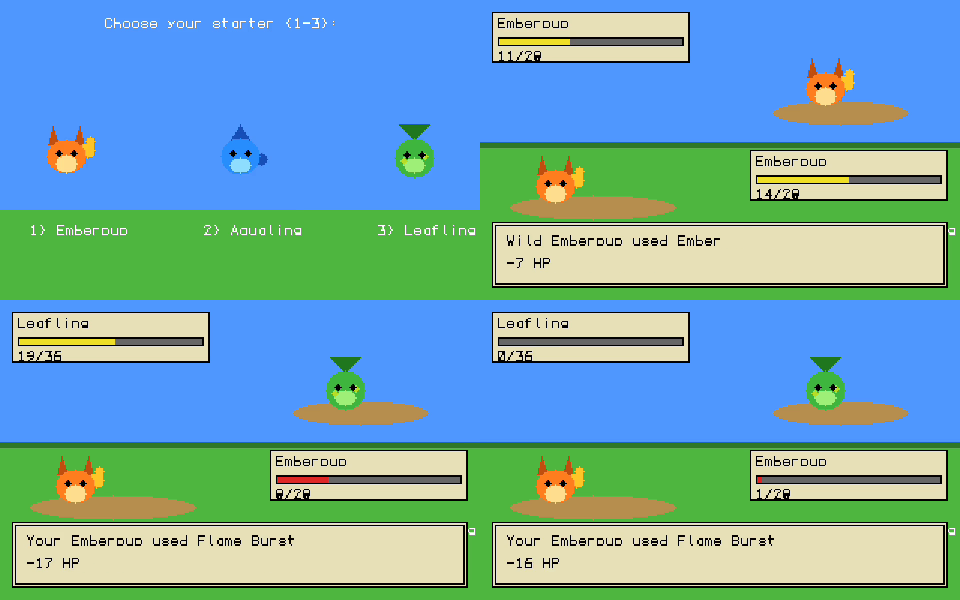}
\caption{\emph{Tiny Monsters} is a video-game written by AI, running in an operating system written by AI, which is compiled by a compiler written by AI. There is not a single line of code written by a human in the whole stack.}
\label{fig:battle}
\end{figure}

\newpage
\section{Introduction}\label{sec:intro}

Power on the machine. A 512-byte boot sector loads a kernel, the kernel mounts its file
system, and a shell prompt appears. Type \texttt{play}, and a
battle screen in the style of a Game Boy Advance lights up.

No human wrote a single line of it: not the boot sector, not the kernel, not the game, and not C
compiler that built all of them. An AI did.

Coding agents are usually judged on changes to code that already works, from single
functions~\cite{chen2021codex} to repository-level issue fixing~\cite{jimenez2024swebench}.
Writing a system from an empty repository is a different task. The agent has to choose
an architecture, keep it coherent over thousands of lines, and debug failures whose only
symptom may be subtle or silent. An operating system is the extreme case.
Below the kernel there is no runtime to report a mistake. A wrong descriptor table entry
resets the machine, and an I/O instruction with the wrong operand size corrupts the disk
without a word. Unix~\cite{ritchie1974unix} and xv6~\cite{cox2006xv6} took extremely skilled people to write.

\begin{table}[t]
\centering
\caption{Every part of \sysname{} is written by AI. The bottom part is the bootstrap: only run once.}
\label{tab:goal}
\begin{adjustbox}{max width=\linewidth}
\begin{tabular}{lll}
\toprule
Part & Written or built by & Claim \\
\midrule
Boot loader: boot sector, second stage & the agent & AI-written kernel \\
Kernel: memory, drivers, CPU exceptions & the agent & AI-written kernel \\
File system and shell & the agent & AI-written kernel \\
Scripting language, compiled language & the agent & AI-written kernel \\
Assembler, C compiler, linker & the agent & AI-written stack \\
Build of the OS & the agent & Full bootstrap \\
Build of the toolchain & the agent  & Full bootstrap \\
\midrule
First binary of the toolchain & the host C compiler huma written & single execution \\
\bottomrule
\end{tabular}
\end{adjustbox}
\end{table}

There are precedents for agents building large systems from nothing. A team of 16 Claude
agents wrote a C compiler that builds Linux, but it relies on GCC's assembler and
linker~\cite{carlini2026ccc}. Hundreds of agents built a web browser in a
week~\cite{lin2026scaling}. A month before this work, Claude Fable 5 wrote the core of an
NT-style kernel in Rust, with a scheduler and a memory manager~\cite{suiche2026ntoskrnl}. All of them are compiled by existing, human-written
compilers (GCC, \texttt{rustc}), so every instruction that runs was chosen by a code
generator a human wrote.

To the best of our knowledge, no one has reported a
bootstrapped operating system written by an AI, in which the AI wrote not only the OS
but also the assembler, compiler and linker that build it.
Even more ambitiously, we want that toolchain compiles
itself to a byte-identical fixed point, and where it is the only toolchain that ever
compiles the OS. This is the contribution of this paper: \sysname{} is a bootstrapped operating system entirely written by AI.

\paragraph{Methodology.} The human author gave one Claude Code agent (\sessionmodel{}) a goal
and an empty repository, and let it work. The human set goals,
answered a few questions from the agent and said ``go''. We study three
artifacts: the code, the git history of the
system\footnote{\url{https://github.com/monperrus/TinyMonOS}, commit \ospin{}.} and
the complete transcript of the
session,\footnote{\url{https://github.com/monperrus/boot-to-claude-paper}} with every
prompt, model response, tool call and tool output. Every number in this paper is computed from these reference sources.

\paragraph{Results.} \sysname{} boots through its own boot loader into a kernel with a
file system, a shell, a scripting language and a compiler to bytecode. Its assembler, C
compiler and linker were written by the agent and compile themselves to byte-identical
binaries. From plan approval, the OS with both languages took \plantoosmin{} minutes;
the toolchain took \tcactivehours{} hours of agent activity. The agent ran its
scripted oracles \noraclesruns{} times and saw \noraclesfailed{} failures, most of them
while switching the build to its own toolchain. The transcript records \ntproduct{}
defects found and fixed in the system, \nttoolchain{} of them in the toolchain; only
\ntreported{} appear in a commit message.

\paragraph{Lessons learned.} Beyond this one build, the experiment yields knowledge that
carries over to agentic work in general:
\begin{itemize}
  \item \textbf{An agent is as good as its oracles.} The agent worked in a loop: change
        the code, run an oracle, iterate until it passes. A defect that its oracle could not
        see stayed in until another oracle did (Section~\ref{sec:oracles}).
  \item \textbf{Most defects are silent.} \ntsilent{} of the \ntproduct{} product defects
        produced no error message at all. An oracle that waits for an error misses them; agentic workflows
        have to state the expected output and compare (Section~\ref{sec:defectlist}).
  \item \textbf{The whole system is the best test program.} Booting the OS caught
        toolchain defects that no unit test covered, because the unit tests cover only what
        the agent thought of, while system tests cover more implicit assumptions.
  \item \textbf{The agent's oracles have defects too.} \ntoracle{} defects were in the
        oracles themselves, and the worst of them made an oracle report success after a
        failure. Oracle design is essential (Section~\ref{sec:guidelines}).
  \item \textbf{The toolchain is harder than the OS.} \nttoolchain{} of the \ntproduct{}
        product defects were in the assembler, compiler and linker. Removing the last
        human-written tool from the build was most of the work.
\end{itemize}

\paragraph{Contributions.}
\begin{itemize}
  \item \textbf{The first bootstrapped operating system written by an AI}, to the best of
        our knowledge. \sysname{} has its own boot loader, kernel, file system, shell and
        two languages, and its own assembler, C compiler and linker, which compile
        themselves to a byte-identical fixed point and are the only tools that build the
        OS. No line of the stack was written by a human.

  \item \textbf{The oracle is what makes it possible.} We show from the
        transcript that the agent heavily used oracles. It ran them
        \noraclesruns{} times, they rejected its work \noraclesfailed{} times, and the
        most telling defects were silent ones that only an oracle booting the whole
        machine caught.
  \item \textbf{A fully transparent open-science package.} We release the system's
        source and history, the complete transcript of the session that built it, with
        every prompt, model response, tool call and tool output. Anyone can
        check the results down to the individual tool call.\footnote{\url{\siteurl}}
  \item \textbf{A game written by an AI, played by an AI}, on a stack written by an AI from the boot sector up.
\end{itemize}

\section{Goal}\label{sec:goal}

\paragraph{Total AI authorship.} We aim at total AI authorship of a computing system: every line of
code, in the system and in everything that builds it, is produced by an AI coding agent.
We use a coding agent to generate this computing system.
We decompose this goal into three claims.
\begin{description}
  \item[Full AI-written kernel.] No line in the kernel is written by
        a human or taken from an existing project.
  \item[Full AI-written build stack.] No line in the build system (compiler, linker) is written by
        a human or taken from an existing project. We do not use a third-party compiler.
  \item[Full bootstrap.] The AI-written build stack builds itself and the kernel. Thus, the system is AI-bootstrapped. An AI-written toolchain (compiler, linker) builds the AI-written OS and builds itself.
\end{description}

The agent in this study is Claude, but this is incidental; the question is whether
total authorship is achievable by an AI agent at all. 

\paragraph{Rationale for the bootstrap.} An OS whose source was written by AI but which \texttt{gcc}
compiles is still a joint work, because every machine instruction that runs was chosen by a code
generator written by humans. The bootstrap removes that dependency. The host compiler is
used once, to produce a first binary of the AI's compiler; from then on the AI's compiler
compiles itself, and the fixed point shows that the result no longer depends on the human-written
compiler. The bootstrap is also a strong oracle in its own right: a
compiler that compiles its own source correctly has to handle a real, large C program, like a kernel.

\paragraph{Two compounding difficulties.} We have two hard problems. The first is
writing an operating system: code that runs on bare hardware, with no runtime underneath, where
a mistake resets the machine or corrupts data without a word. The second is writing a compiler
strong enough to compile both that operating system and itself. 
Building a C compiler is
hard even with the abundant resources of Anthropic: their team of 16 agents spent about \$20{,}000
on a C compiler that builds Linux, yet, they still relies on GCC's assembler and
linker~\cite{carlini2026ccc}. 

The two difficulties
compound. When the OS misbehaves after bootstrap, the cause may be in the OS or in the compiler that
built it, and the symptom cannot tell them apart. An \texttt{in} instruction encoded without its
operand-size prefix, for example, showed up as a corrupted file system
(see Section~\ref{sec:defectlist}).

\paragraph{Scope.} The target is a complete, if small, computing system: one that boots
on a bare machine and that lets a user write and run programs in
languages of its own. Concretely, a boot loader, a kernel, a file system, an interactive
shell,  and a compiled language. Then an assembler, a C compiler and a linker, the tools that build all of it.
Table~\ref{tab:goal} sums up what is required of each part.

\section{\sysname}\label{sec:system}

\sysname is the entirely AI-written kernel, compiled by an entirely AI-written compiler.

\subsection{Kernel}

\paragraph{Boot and kernel.} The BIOS loads a 512-byte boot sector, which loads a second
stage. The second stage enables the A20 line, loads the kernel, installs a flat GDT and
switches to 32-bit protected mode. The kernel owns its GDT and an IDT that covers CPU
exceptions only. It has a free-list allocator over a fixed 1\,MB heap, VGA text and
COM1 serial output, a polled ATA PIO disk driver, and a polled PS/2 keyboard driver.

\paragraph{Deliberately single-tasking.} The kernel never enables interrupts after boot.
It has no scheduler, no paging and no user mode. The shell and both language runtimes run
in kernel context.

\paragraph{Above the kernel.} \emph{tinyfs} is a flat file system with a superblock, a
free-block bitmap, a fixed inode table and 4\,KB maximum file size. Its format code is
compiled twice, once into the kernel and once into the host tool that formats the disk
image. The shell offers
\texttt{ls}, \texttt{cat}, \texttt{write}, \texttt{rm}, \texttt{run}, \texttt{build} and
\texttt{exec}. \texttt{run} executes a dynamically typed script with a tree-walking
interpreter. \texttt{build} compiles a statically scoped, integer-only C-like language to
stack bytecode, and \texttt{exec} runs it in a VM that supports recursion. The shell's
\texttt{play} command starts a game, described in Section~\ref{sec:game}.

\paragraph{The build toolchain.} Three tools replace \texttt{nasm}, \texttt{gcc},
\texttt{ld} and \texttt{objcopy}:
\begin{itemize}
  \item an \emph{assembler} for a subset of NASM syntax that the OS's assembly files
        and the compiler's output use, with no macro preprocessor. It writes either a
        flat binary (the boot sectors) or an ELF32 relocatable object;
  \item a \emph{C compiler} for tinyC, a C dialect scoped to what the code base needs:
        \texttt{\#include} and constant \texttt{\#define} only, structs, unions, enums
        and function pointers, no type wider than 32 bits. It emits assembly text for the
        assembler, either freestanding for the kernel or as a static Linux program that
        calls the kernel directly through system calls;
  \item a \emph{linker} that merges ELF32 objects, applies the two x86 relocation types
        the assembler produces, and writes either the flat kernel image or a static
        Linux executable. Writing the flat image itself made \texttt{objcopy}
        unnecessary.
\end{itemize}
A minimal C library (unbuffered \texttt{FILE} I/O, a bump allocator, a subset of
\texttt{printf}) lets the toolchain's own source, which uses ordinary libc calls, be
compiled by the toolchain. Table~\ref{tab:size} gives the size of each part.

\paragraph{Bootstrap path.} The toolchain is written in C, so its very first binary has
to come from a compiler that already exists. The host C compiler (\texttt{gcc} by
default) is used exactly once, to build Stage~1 of the
toolchain from its source. Stage~1 then compiles the same source into Stage~2, and
Stage~2 compiles it again into Stage~3.  The emulator, \texttt{qemu-system-i386}, stays external: it is the
machine \sysname{} runs on, not part of what builds it.

\paragraph{Fixed point.} Let $S_i$ be the binaries of stage $i$. The
bootstrap is accepted only when
\begin{equation}\label{eq:fixpoint}
  S_2 = S_3 \quad \text{byte for byte, for each of } \texttt{as},\ \texttt{cc},\ \texttt{ld}.
\end{equation}
Equation~\eqref{eq:fixpoint} shows that the compiler compiles itself to the same result
it was compiled to, so the bytes of Stage~2 come from the toolchain's own code
generator, not from the host compiler's. Stage~2 then
builds the boot loader, the kernel and the disk-image tool; the host compiler never
compiles any of them. 

\begin{table}[t]
\centering
\caption{Size of \sysname{} at commit \texttt{a4a64cb}, in physical lines including comments.}
\label{tab:size}
\begin{adjustbox}{max width=\linewidth}
\begin{tabular}{lr}
\toprule
Component & Lines \\
\midrule
Boot loader (\texttt{boot/}) & 218 \\
Kernel (\texttt{kernel/}) & 5003 \\
\quad of which languages (\texttt{kernel/lang/}) & 2041 \\
\quad of which game (\texttt{kernel/game/}) & 748 \\
\quad of which tinyfs (\texttt{kernel/fs/}) & 333 \\
\quad of which shell (\texttt{kernel/shell/}) & 286 \\
Toolchain (\texttt{toolchain/}) & 6430 \\
\quad of which C compiler & 3109 \\
\quad of which assembler & 2050 \\
\quad of which linker & 816 \\
Build and test tools (\texttt{tools/}) & 682 \\
Test fixtures (\texttt{tests/}) & 424 \\
Design documents (\texttt{docs/}) & 1099 \\
\bottomrule
\end{tabular}
\end{adjustbox}
\end{table}

\subsection{Tiny Monsters}\label{sec:game}

Now that we have a working operating system,  we implement and play with a Pokemon like game.
The result, \emph{Tiny Monsters}, is the most visible demonstration of the full AI-written
stack. Every pixel in Figure~\ref{fig:battle} is drawn by code Claude wrote, compiled by
the compiler Claude wrote, on a kernel Claude wrote.

\paragraph{Rules.} Three species, Emberpup, Aqualing and Leafling, one per type in a
rock-paper-scissors cycle: Fire beats Grass, Grass beats Water, Water beats Fire. A move
with the type advantage does 1.5 times its damage, one against it half. The player picks
a starter and meets wild monsters whose HP grows with each win. HP does not regenerate,
and five wins make the player Champion.

\paragraph{Made for a machine player.} Until then the kernel only read the keyboard; the
serial port was output only. The agent added serial input and made the shell and the
game poll both sources (\inputlines{} lines in \texttt{input.c}), so a human at the
keyboard and a program on the serial line can drive the same session interchangeably.
The game speaks a plain text protocol: it prints the screen as text, and reads a digit,
\texttt{run} or \texttt{quit}. Playing found a \defect{D-C09-01}{real bug}: the random move
picker took a modulo of a 32-bit value with the sign bit set, which \sysname{}'s compiler
only supports as signed division, and read a garbage move name with absurd damage.

\paragraph{A graphical battle screen.} Asked next for \emph{``a good looking GUI like
Pokemon Ruby on GBA''}, the agent wrote a driver for VGA mode 13h (320$\times$200, 256
colors) that programs the video card's registers directly, without BIOS calls
(\vgadriver{} lines), and a bitmap font of its own design (\fontlines{} lines). The three
monsters are not images: \texttt{sprites.c} draws them from filled circles, ellipses and
triangles in \gamesprites{} lines. The battle scene, with sky, ground, platforms, HP bars
and a dialogue box, takes \gamegui{} lines on top of the \gamerules{} lines of game
logic. The graphics are purely additive. The text sent over serial is byte for byte the
same, so the machine player sees no difference. Returning to the shell restores text
mode, which took a debugging session of its own because \defect{D-C09-05}{mode 13h
overwrites the video memory plane} where the text font lives.

\paragraph{Built by AI, played by AI.} The game was designed for an AI player from the first
request (\emph{``a Pokemon like game such that an AI agent later plays over serial port''}),
and its first player was Claude itself. Before the human saw the game, the serial game
test that the agent wrote along with it had already fought full battles over the serial
port (Table~\ref{tab:oracles}). When asked to play for five minutes, the agent booted the
OS , wrote a player that connects to QEMU's serial socket, and played
\playruns{} runs in \playseconds{} seconds, restarting after each faint, with no human and no
keyboard in the loop.

\subsection{Process Metrics}\label{sec:history}

The work happened in one Claude Code session using \sessionmodel{}, with
\nassistant{} model responses and \ntoolcalls{} tool calls, of which \nbash{} shell
commands and \nedit{} file writes or edits. The context window filled up and was
summarized \ncompactions{} times. Table~\ref{tab:phases}, in the appendix, splits the session at each phase commit. Of the \ncommits{} commits in the repository, all but two were committed by
the agent. The other two are the initial empty commit and a ten-line Makefile committed (not written) by the human, by habit after \texttt{git status}.

\paragraph{The toolchain over four days.} Replacing the host tools took
\tcactivehours{} hours of agent activity, about nine times the OS. The assembler had to
grow from the instructions in four hand-written \texttt{.S} files to everything the C
compiler emits. The kernel source was first rewritten into the tinyC dialect while still
building with \texttt{gcc} (phase T3), so that each change could be checked as
behavior-preserving before the new toolchain was in the loop.

\paragraph{The fixed point.} Phase T8 implemented the bootstrap path and reached
Equation~\eqref{eq:fixpoint}. Compiling the toolchain with itself required features it
had never needed for the kernel: variadic functions, multi-dimensional arrays, and the
minimal C library.

\section{Defects}\label{sec:defectlist}

This section describes the defects the agent made and how they came to light;
Section~\ref{sec:oracles} then describes the oracles that revealed them.

\subsection{Method.} Commit messages report the defects the agent chose to mention. To
see all of them, we cut the transcript into ten chunks at phase boundaries and had ten
annotator agents list every defect the agent encountered from commit analysis, with what was wrong, the tool call
whose output first showed it, the edits that fixed it, the check that confirmed the fix,
and the commit message that reports it, if any.

Each quotes its evidence
verbatim. A script then checked every cited tool call against the transcript and every
quote against the chunk. 
\href{\siteurl/defects.html}{All records} are on
the verification site, each with its type (product, oracle tooling, environment), its
component, the quote, and links to the tool calls that showed, fixed and
confirmed it.

\subsection{Results.} The transcript holds \ntdefects{} defects: \ntproduct{} in
\sysname{} itself, \ntoracle{} in the oracle tooling, and \ntenv{} in the environment
(Table~\ref{tab:defects}).

\ntreported{} of the \ntproduct{} product defects are
reported in a commit message; the other \ntunreported{} were found and fixed between
commits and left no trace in the commit history, they are visible only in the trajectory. The toolchain accounts for \nttoolchain{} of
the \ntproduct{}, the C compiler alone for \ntcc{}: building a compiler strong enough
was the harder half of the task. \ntmultiattempt{} defects took more than one failing
check to fix, up to \ntmaxattempts{}. 

\begin{table}[t]
\centering
\caption{Defects recovered from the full transcript, by component and by what revealed them. \emph{In commit msg.} counts those that a commit message also reports. \emph{Boot oracles} are the OS system test, the boot self-test, the CPU exception handlers and the bootstrap comparison.}
\label{tab:defects}
\begin{adjustbox}{max width=\linewidth}
\begin{tabular}{lrrrrrrrr}
\toprule
Component & Defects & In commit msg. & Build & Toolchain tests & Boot oracles & Manual run & Screenshot & Code reading \\
\midrule
Boot loader and kernel & \href{\siteurl/defects.html\#row-0}{7} & 2 & 1 &  & 1 &  &  & 5 \\
Languages (script, compiler, VM) & \href{\siteurl/defects.html\#row-1}{1} & 1 &  &  & 1 &  &  &  \\
Game and VGA & \href{\siteurl/defects.html\#row-2}{8} & 5 &  &  & 1 & 2 & 4 & 1 \\
Assembler (AS) & \href{\siteurl/defects.html\#row-3}{27} & 16 & 10 & 2 & 7 &  &  & 8 \\
C compiler (CC) & \href{\siteurl/defects.html\#row-4}{57} & 30 & 23 & 9 & 8 &  &  & 17 \\
Linker and C library (LD, hostlib) & \href{\siteurl/defects.html\#row-5}{7} & 3 & 4 &  &  & 2 &  & 1 \\
Build rules & \href{\siteurl/defects.html\#row-6}{6} & 2 & 2 &  &  & 2 &  & 2 \\
\midrule
\textbf{Product total} & \href{\siteurl/defects.html}{113} & 59 & 40 & 11 & 18 & 6 & 4 & 34 \\
\midrule
Oracle tooling & \href{\siteurl/defects.html\#oracle}{12} & 1 & 2 & 2 &  & 5 & 1 & 2 \\
Environment & \href{\siteurl/defects.html\#environment}{2} & 0 &  &  &  & 2 &  &  \\
\bottomrule
\end{tabular}
\end{adjustbox}
\end{table}

\paragraph{What reveals defects.} \ntbuildrevealed{} product defects showed up at build
time, mostly the compiler rejecting C it did not yet support. \ntbootrevealed{} were
caught by an oracle that boots the machine, and \ntsilent{} had a silent symptom: wrong
output, a hang, corrupted data or a wrong picture, with no error message. \ntcodereading{}
were found by the agent reading its own code, typically while investigating a nearby
failure. Three patterns show why the end-to-end oracles mattered.

\paragraph{Many defects are silent.} Silent means that the defective component reported
nothing: no compiler error, no assembler warning, no CPU exception. The machine kept
running, and the defect showed up later, elsewhere, as wrong behavior. An
\defect{D-C06-04}{\texttt{in}/\texttt{out} instruction} encoded without its operand-size
prefix read two 16-bit words per access instead of one, which surfaced as file system
superblock corruption. The compiler's lexer had \defect{D-C06-13}{no case for the
\texttt{\textbackslash b} escape} and read the backspace character literal as the letter
\texttt{b}, so every typed \texttt{b} in a shell command acted as a backspace. In the
scripting interpreter, \defect{D-C01-08}{integer addition} fell through to a default
branch returning 0, so \texttt{i = i + 1} looped forever. All three were caught by the OS
system test, because it compares what the machine prints with what it should print: the
self-test reported failing file system checks, the echo of a typed \texttt{build} came back
as \texttt{uild}, and the test timed out waiting for the line \texttt{x = 14}. The
exception handlers, which catch crashes, saw none of them. The take-away is that an oracle that
only waits for an error message misses most defects; it must state the expected output and
compare.

\paragraph{Defects hide behind other defects.} A \defect{D-C04-12}{three-operand
\texttt{imul}} that dropped its immediate went unnoticed because code that wrote and read arrays used the same wrong
index formula, so the errors cancelled. It surfaced only when a host program built by the agent's
compiler, assembler and linker read \texttt{argv}, an array written by the Linux kernel, and
failed to open its input file. The take-away is that a test in which the same code writes and
reads the data checks that the code agrees with itself, not that it is correct. Such
defects show only where the data crosses into code the agent did not write, or is checked
against a value computed independently.

\paragraph{System tests catch what unit tests miss.} The assembler's
\defect{D-C08-08}{\texttt{.dq} directive} computed its high 32 bits with a shift by 32, which x86 masks to a
shift by 0. No toolchain test used \texttt{.dq}. The boot loader's disk address packet
does, and only the OS system test caught it. The machine never got past the boot sector.
The take-away is that unit tests cover what their author thought of, and here the author is the
agent that wrote the bug. The real product exercises what it actually uses. Fast unit
tests do not replace the end-to-end test; they make it the last line, not the only one.

\subsection{Defects in the oracle loop}\label{sec:oracledefects}

The oracles are code too and \ntoracle{} of the defects were
in them. They differ in nature from product defects. A product defect makes the oracle
say no, while an oracle defect makes it say the wrong thing, or nothing. Three kinds
appear.

\paragraph{Oracles that could not fail.} The toolchain test runner
\defect{D-C03-18}{had no expected exit code} for one fixture, so that test was silently
skipped. The \defect{D-C04-13}{pointer test kept passing} with the broken three-operand
\texttt{imul}, because it wrote and read arrays with the same wrong index formula. Worse,
\defect{D-C04-17}{ad-hoc verification commands} printed success lines such as
\texttt{IDENTICAL} after a step had already failed, and misled the diagnosis. A
false \texttt{PASS} is the most dangerous output an oracle can produce, because the agent
trusts it. None of the three was caught by an oracle. The skipped fixture was visible
only because the runner printed a \texttt{SKIP} line, and the agent registered the
missing expected value. The other two were found by the agent reading the output against
itself: an error line directly followed by a success line. For the ad-hoc commands, the
agent reran each step on its own with its exit code printed. The weak pointer test,
found while chasing the \texttt{imul} defect, was never strengthened.

\paragraph{Oracles that could not observe.} A \defect{D-C09-03}{QEMU serial socket} in
\texttt{server,nowait} mode discards everything the guest prints before a client
connects, so a script waiting for the boot banner timed out on a healthy machine. A
\defect{D-C10-06}{screencast recorder} assumed 320$\times$200 frames where QEMU produces 640$\times$400, and
the agent first blamed a torn file read. Both were caught by their symptom only: the
script waited for a banner that a healthy machine had already printed, and the frames,
looked at by the agent, were sheared and double-exposed. Neither symptom points at the
instrument, so the agent first suspected the system or the wrong part of the
instrument. A stricter re-read, its first fix for the recorder, rejected every frame and
produced an empty video. Each time, the agent had to debug its instrument before it could
trust it again.

\clearpage
\section{Oracles for Coding Agents}\label{sec:oracles}

\begin{table}[t]
\caption{The oracles of \sysname{}. \emph{Runs} counts the times the agent executed the oracle during the session, each run linked to the list of runs on the verification site; \emph{No verdict} means the output carried neither verdict, typically a hang killed by a timeout, a crash of the tool under test, or a build error before the test started.}
\label{tab:oracles}
\begin{adjustbox}{max width=\linewidth}
\begin{tabular}{p{2.6cm}p{6cm}p{3cm}rrr}
\toprule
Oracle & What it checks & Verdict & Runs & Failed & No verdict \\
\midrule
CPU exception handlers & any CPU fault, e.g.\ an invalid opcode from a mis-encoded instruction & \texttt{EXCEPTION vector=$n$}, then halt & \multicolumn{3}{c}{in every boot} \\
Boot self-test & 19 checks: string functions, allocator with coalescing, tinyfs create/write/read/delete & \texttt{SELFTEST PASS} or \texttt{FAIL} & \multicolumn{3}{c}{in every boot} \\
OS system test & boots headless QEMU, types shell commands through the QEMU monitor, compares serial output with 2 expected outputs; both languages and the game menu & \texttt{run\_tests: PASS} & \href{\siteurl/oracles.html\#os}{51} & 15 & 4 \\
Serial game test & plays a full battle over the serial port alone, then checks the shell answers & \texttt{serial\_game\_\allowbreak{}test: PASS} & \href{\siteurl/oracles.html\#serial}{8} & 0 & 1 \\
Toolchain tests & 16 C programs compiled by CC, AS and LD, run on the host; exit code compared with the expected value & \texttt{run\_toolchain\_\allowbreak{}tests: PASS} & \href{\siteurl/oracles.html\#toolchain}{64} & 8 & 5 \\
Bootstrap fixed point & Stage~2 and Stage~3 of AS, CC and LD are byte-identical & \texttt{bootstrap: PASS} & \href{\siteurl/oracles.html\#bootstrap}{2} & 0 & 0 \\
Reference tools & while \texttt{nasm}, \texttt{gcc}, \texttt{ld} and \texttt{objcopy} are still in the build: a change must pass the tests with the old tools first; LD's flat output must equal \texttt{objcopy}'s byte for byte & same as the tests above & \multicolumn{3}{c}{---} \\
Boot budget & the kernel image fits the sectors the boot loader reads & build error & \multicolumn{3}{c}{in every build} \\
Screenshots & the graphical battle screen, from QEMU frame-buffer dumps & the agent looks at the image & \multicolumn{3}{c}{25 images read} \\
\bottomrule
\end{tabular}
\end{adjustbox}
\end{table}

A coding agent is as strong as the oracles it uses. The agent proposes a change, the oracle
judges it, and the agent iterates until the oracle agrees. The outcome quality depends on the oracle. A weak one lets defects through, and a missing
one leaves the agent guessing.
Oracles
serve twice in the loop. They reveal a defect, and they confirm its fix.

\sysname{} had strong oracles, listed in
Table~\ref{tab:oracles}, and the agent consulted them constantly: it ran the four
scripted ones \noraclesruns{} times, and \noraclesfailed{} of those runs failed. 
\href{\siteurl/oracles.html}{Every run} is listed on the verification site with the
defects it revealed and the fixes it confirmed, and every defect links back to the oracle
that found it: \ntfirstinrun{} of the \ntdefects{} defects were first seen in a scripted
oracle run, and \ntconfirmedinrun{} fixes were confirmed by one.

\subsection{Kinds of oracle}\label{sec:oraclekinds}

The oracles of Table~\ref{tab:oracles} decide correctness in three different ways, which
differ in what they need to exist beforehand.

\paragraph{Expected-output oracle.} The test states what the system must produce for a
given input, and compares. The OS system test compares the serial console with the
expected text of each shell session, the toolchain tests compare the exit code of each
compiled program with a registered value, and the boot self-test checks each result
against an expected value. It needs someone, here the agent, to know the right answer in
advance, and it checks only what that someone thought of. This is the agentic version of test-first development.

\paragraph{Differential oracle.} The same input goes through the system and through a
reference implementation, and the two outputs must agree. During the transition the
old tools were still available, so the kernel was first rewritten into the tinyC dialect
while still building with \texttt{gcc}, so the OS test could confirm each rewrite before
the new compiler was in the loop, and the linker's flat output was compared byte for
byte with \texttt{objcopy}'s. It needs no expected value, but it needs a reference. With differential oracles, the ultimate goal is to remove the references from the build.

\paragraph{Fixed-point oracle.} The system is applied to itself, and the result must not
change: the toolchain compiled by Stage~1 and by Stage~2 must be byte-identical
(Equation~\eqref{eq:fixpoint}). It needs neither an expected value nor a reference, which
is why it took over once the old tools were gone.

\subsection{How the oracles observe and decide}\label{sec:oracleobserve}

\paragraph{Single observation channel.} Every expected-output oracle that runs the OS
compares the same thing, the text on the serial console. For this, the kernel writes every line of
output to the COM1 serial port. Even a crash is written, because
the CPU exception handlers print the number of the faulting exception to the serial port
before halting the machine. The oracles run QEMU without a display and save everything
the serial port receives to a file, which the agent reads. With this design, a crash, a
hang and a wrong answer all are text the agent can compare with the expected output. The
agent built this agent-native observation channel in phase 0, before there was anything to test, and kept it for
six days.

\paragraph{Binary verdicts.} Each scripted oracle, of whichever kind, ends with one line,
\texttt{PASS} or \texttt{FAIL}: the expected-output comparison matched or not, the
differential comparison agreed or not, the fixed point was reached or not. There is
no room for interpretation. When a run produces neither, because the machine hung until the
timeout, the tool under test crashed, or the build broke first, the absence of
\texttt{PASS} is itself the failure.

\subsection{How the oracles combine}\label{sec:oraclecombine}

\paragraph{Combination over time.} The three oracle kinds took turns over the session.
Expected-output oracles ran throughout. Differential oracles ran while the old tools were
still in the build, until the build switched over to the new toolchain (phase T7). The
fixed-point oracle took over in phase T8, when the bootstrap removed the last reference.
Switching the build over is where the oracles mattered most: \ntsevenfailed{} of the
\noraclesfailed{} failing oracle runs of the whole session happened in phase T7 alone.

\paragraph{End to end.} The OS system test is an expected-output oracle on the whole
system. It boots the boot loader and kernel built by the toolchain under test, and drives
them as a user would. Any defect anywhere on that path, in the assembler's encoding, the
compiler's code generator or the linker's relocations, shows up as a wrong line on the
serial console. It also covers the gap of the fixed-point oracle. A toolchain that
reaches its fixed point while miscompiling something the OS uses still fails the OS
system test. This is what made the toolchain safe to replace, because the OS itself is the
largest test program the toolchain can be checked on.

\paragraph{Fast and slow expected-output oracles.} The toolchain tests and the OS system
test are expected-output oracles at two scales. The toolchain tests run
\ntoolchainfixtures{} small programs on the host in seconds; the OS test boots a machine.
The agent ran them \ntoolchaintestruns{} and \nostestruns{} times respectively, iterating
on the fast oracle and confirming on the slow one.

\paragraph{Regression tests.} When a failure of the slow oracle exposed a defect, the
agent added a regression test for its class, turning the defect into one more expected
output of the fast oracle.

\subsection{Guidelines for Oracle Design}\label{sec:guidelines}

The defects of Section~\ref{sec:defectlist}, in the product and in the oracles, and the
oracles of Section~\ref{sec:oracles} suggest the following guidelines
for designing oracles for coding agents. 

\begin{enumerate}
  \item \textbf{Combine all kinds of oracle: expected output, differential and fixed-point} (see Section~\ref{sec:oraclekinds}). An
        oracle that only waits for an error misses silent defects. State the expected
        output and compare. Code that writes and reads its own data agrees with itself
        even when wrong. Check it against a reference implementation while one exists,
        against code the agent did not write (the Linux kernel, the BIOS), or, when possible, against itself with a fixed point.
  \item \textbf{Instruct the agent to write and run system tests, end to end.} They are
        excellent at finding failures. The unit tests cover what the agent thought of;
        the whole system exercises what it actually uses. Here the test that boots the
        OS caught defects that no toolchain test covered (Section~\ref{sec:defectlist}).
  \item \textbf{Derive verdicts robustly.} A step that
        prints a success line unconditionally or match the wrong output lies after a failure. A missing verdict, a
        hang or a crash of the harness, counts as a failure
        (Table~\ref{tab:oracles}, \emph{No verdict}).
  \item \textbf{Validate the oracle on a healthy system.} When a check fails, first
        ask whether the instrument can observe at all. Capture from the start, check frame
        sizes, and run it once on a known-good build.
  \item \textbf{Keep the test harness out of its own way.} By test harness we mean the
        scripts and commands the agent writes to start, drive and stop the system under
        test, here QEMU, not the coding agent's harness. A
        cleanup that matches processes by a pattern can match that shell and kill the
        test run with it, which looks exactly like a broken system. Instruct the agent to clean up by process
        id, and to keep sockets, timeouts and cleanup in the test harness, never mixed
        with the system under test.
\end{enumerate}

\section{Threats to Validity}\label{sec:threats}

The defect analysis relies on annotator agents reading the transcript. They may have
missed defects or misjudged a cause; we mitigated this with verified verbatim
evidence, an automatic check of every quote and tool call, a coverage check against
the commit messages, and a manual check of a random sample. 

Oracle verdicts are read from the transcript by matching each
script's \texttt{PASS} and \texttt{FAIL} lines. A run counts only when a shell command
executes the oracle's script; a single test case run by hand does not, so the run counts
are a lower bound.

Active time is
estimated from transcript timestamps, counting gaps shorter than ten minutes; a long
build without output would be counted as idle. This is one session with one model, so
it shows that the result is possible, not how often it happens.

\section{Related Work}\label{sec:related}

\paragraph{Complex systems built by coding agents.} Since 2025, practitioners have
reported coding agents building systems far beyond single features, mostly in blog posts. The closest to \sysname{} is Anthropic's C compiler~\cite{carlini2026ccc}:
16 parallel Claude Opus 4.6 agents, nearly 2{,}000 sessions and about \$20{,}000 produced a
100{,}000-line Rust compiler that builds Linux 6.9 for three architectures. It has no
assembler or linker of its own and calls GCC for 16-bit x86 code.
\sysname{} has it own AI-written assembler, compiler and linker and reaches a self-hosting fixed point. 

Cursor ran hundreds of
agents for about a week in a planner, worker and judge hierarchy to build FastRender, a
web browser of over a million lines~\cite{lin2026scaling}. Huntley ran Claude in a loop
with a single prompt for three months and obtained \emph{cursed}, a programming language
with an LLVM back end~\cite{huntley2025cursed}. Stenström built JustHTML, an HTML5 parser
that passes the 9{,}200 tests of html5lib-tests, and reports the lesson this paper also
draws: a complete test suite is what lets the agent measure its own
progress~\cite{stenstrom2025justhtml}. All three systems are turned into running code by
tools written by humans, such as LLVM for \emph{cursed} and the Python interpreter for
JustHTML; \sysname{} is the only one in which the agent also wrote, and bootstrapped, the
tools that build it. 

Closest in domain, Suiche
reports that Claude Fable 5 wrote the core of \texttt{ntoskrnl-rs}, an NT-style x86-64
kernel in Rust that boots in QEMU and passes its self-tests. ~\cite{suiche2026ntoskrnl}. As a kernel it is more ambitious than
\sysname{}'s: it has a scheduler, a memory manager with page tables, and interrupt-driven
timers, and later work ran unmodified Windows drivers and programs. But it is a Rust
project compiled by \texttt{rustc}, and its core has no file system, shell or user
programs. What set \sysname{} apart is a whole stack from boot sector to a
self-hosted toolchain with no third-party compiler left in the build.

\paragraph{Benchmarks for building from scratch.} HumanEval~\cite{chen2021codex} measures
function-level synthesis, and SWE-bench~\cite{jimenez2024swebench} measures issue
resolution in existing repositories. Commit0~\cite{zhao2025commit0} asks agents to
implement 54 Python libraries from a specification and unit tests. ProjectEval~\cite{liu2025projecteval}
and ProjDevBench~\cite{lu2026projdevbench} evaluate whole projects generated from natural
language requirements. In all of them the oracle is given to the agent. In \sysname{} the
agent wrote its oracles too, as it would be done in the real world (Section~\ref{sec:oracles}).

\paragraph{Agents that play games.} In Claude Plays Pokémon, Claude plays Pokémon Red
through an emulator, reading the screen and pressing buttons~\cite{techcrunch2025pokemon}.
Tiny Monsters reverses the setup. The agent wrote the game, the OS under it and the
toolchain that built both, and designed a text interface over the serial port so that an
agent can play without reading pixels.

\section{Conclusion: Built on Human Code}\label{sec:conclusion}

A language model wrote a small, single-tasking operating system, together with the
assembler, compiler and linker that build it, up to a byte-identical self-hosting fixed
point. Every line of \sysname{} was written by an AI,
yet the system is founded on human code, and will remain so forever, in two ways.

\paragraph{The model learned from human code.} AI could write a boot sector, a file
system and a C compiler because it was trained on decades of code that people wrote,
documented and shared: operating systems, compilers, assemblers, textbooks, manuals and
answers to questions. None of \sysname{} could have been written without that body of
human work.

\paragraph{The bootstrap starts from a human compiler.} The toolchain compiles itself, but
its very first binary, Stage~1, was produced by \texttt{gcc}, the host C compiler, a
program written by people over decades. The fixed point of Equation~\eqref{eq:fixpoint}
shows that the bytes of Stage~2 no longer depend on \texttt{gcc}'s code generator; it does
not erase where they came from. As Thompson observed, a self-hosted toolchain carries its
first generation with it~\cite{thompson1984trusting}. And underneath both, we also have human intelligence that the agent took as given: the x86 instruction set, the BIOS boot protocol, the
ELF format, the C language.

An AI-written system is therefore not a break from human computing but its continuation.
The collective effort of the people who wrote, published and maintained the code the
models learned from, and the compilers every bootstrap starts from, is an essential and
eternal contribution of humanity to computing.

\bibliographystyle{abbrvnat}
\bibliography{refs}

\begin{thebibliography}{14}
\providecommand{\natexlab}[1]{#1}
\providecommand{\url}[1]{\texttt{#1}}
\expandafter\ifx\csname urlstyle\endcsname\relax
  \providecommand{\doi}[1]{doi: #1}\else
  \providecommand{\doi}{doi: \begingroup \urlstyle{rm}\Url}\fi

\bibitem[Carlini(2026)]{carlini2026ccc}
N.~Carlini.
\newblock Building a {C} compiler with a team of parallel {Claudes}.
\newblock Anthropic Engineering blog,
  \url{https://www.anthropic.com/engineering/building-c-compiler}, Feb. 2026.

\bibitem[Chen et~al.(2021)Chen, Tworek, Jun, Yuan, et~al.]{chen2021codex}
M.~Chen, J.~Tworek, H.~Jun, Q.~Yuan, et~al.
\newblock Evaluating large language models trained on code.
\newblock arXiv:2107.03374, 2021.

\bibitem[Cox et~al.(2006)Cox, Kaashoek, and Morris]{cox2006xv6}
R.~Cox, M.~F. Kaashoek, and R.~Morris.
\newblock xv6: a simple, {Unix}-like teaching operating system.
\newblock \url{https://pdos.csail.mit.edu/6.828/2023/xv6.html}, 2006.

\bibitem[Huntley(2025)]{huntley2025cursed}
G.~Huntley.
\newblock I ran {Claude} in a loop for three months, and it created a {GenZ}
  programming language called cursed.
\newblock \url{https://ghuntley.com/cursed/}, Sept. 2025.

\bibitem[Jimenez et~al.(2024)Jimenez, Yang, Wettig, Yao, Pei, Press, and
  Narasimhan]{jimenez2024swebench}
C.~E. Jimenez, J.~Yang, A.~Wettig, S.~Yao, K.~Pei, O.~Press, and K.~Narasimhan.
\newblock {SWE}-bench: Can language models resolve real-world {GitHub} issues?
\newblock In \emph{International Conference on Learning Representations
  (ICLR)}, 2024.

\bibitem[Lin(2026)]{lin2026scaling}
W.~Lin.
\newblock Scaling long-running autonomous coding.
\newblock Cursor blog, \url{https://cursor.com/blog/scaling-agents}, Jan. 2026.

\bibitem[Liu et~al.(2025)]{liu2025projecteval}
K.~Liu et~al.
\newblock {ProjectEval}: A benchmark for programming agents automated
  evaluation on project-level code generation.
\newblock In \emph{Findings of the Association for Computational Linguistics
  (ACL)}, 2025.

\bibitem[Lu et~al.(2026)]{lu2026projdevbench}
P.~Lu et~al.
\newblock {ProjDevBench}: Benchmarking {AI} coding agents on end-to-end project
  development.
\newblock arXiv:2602.01655, 2026.

\bibitem[Ritchie and Thompson(1974)]{ritchie1974unix}
D.~M. Ritchie and K.~Thompson.
\newblock The {UNIX} time-sharing system.
\newblock \emph{Communications of the ACM}, 17\penalty0 (7):\penalty0 365--375,
  1974.

\bibitem[Stenstr{\"o}m(2025)]{stenstrom2025justhtml}
E.~Stenstr{\"o}m.
\newblock How {I} wrote {JustHTML} using coding agents.
\newblock
  \url{https://friendlybit.com/python/writing-justhtml-with-coding-agents/},
  Dec. 2025.

\bibitem[Suiche(2026)]{suiche2026ntoskrnl}
M.~Suiche.
\newblock {Fable 5} wrote a {Windows} kernel in 38 minutes.
\newblock \url{https://www.msuiche.com/posts/fable-5-windows-kernel/}, June
  2026.

\bibitem[{TechCrunch}(2025)]{techcrunch2025pokemon}
{TechCrunch}.
\newblock {Anthropic's Claude AI} is playing {Pok\'emon} on {Twitch}, slowly.
\newblock
  \url{https://techcrunch.com/2025/02/25/anthropics-claude-ai-is-playing-pokemon-on-twitch-slowly},
  Feb. 2025.

\bibitem[Thompson(1984)]{thompson1984trusting}
K.~Thompson.
\newblock Reflections on trusting trust.
\newblock \emph{Communications of the ACM}, 27\penalty0 (8):\penalty0 761--763,
  1984.

\bibitem[Zhao et~al.(2025)Zhao, Jiang, Lee, Chiu, Cardie, Gall{\'e}, and
  Rush]{zhao2025commit0}
W.~Zhao, N.~Jiang, C.~Lee, J.~T. Chiu, C.~Cardie, M.~Gall{\'e}, and A.~M. Rush.
\newblock Commit0: Library generation from scratch.
\newblock In \emph{International Conference on Learning Representations
  (ICLR)}, 2025.

\end{thebibliography}

\appendix

\section*{Appendix: What the Human Did}\label{sec:human}

Table~\ref{tab:prompts} lists the human's prompts verbatim, up to the request to play the
game. Three kinds stand out.

\paragraph{Direction.} Five prompts set the direction, quoted here verbatim:
\begin{itemize}
  \item \emph{``I want to ask Claude to write a tiny OS with a tiny file system, with a tiny
        compiler and scripting language, all to be used in qemu/kvm or equivalent. the goal
        is that 100\% of lines of code come from Claude. Make a plan.''}
  \item \emph{``now make a plan to implement AS CC HOSTCC LD OBJCOPY and make such that we
        can build the system with generated tools only. The CC should compile the other
        tools and be self-hosted of course.''}
  \item \emph{``now install and test the self-hosted compiler inside tinyos.''}
  \item \emph{``implement in tinyos a Pokemon like game such that an AI agent later plays
        over serial port.''}
  \item \emph{``very good, it works, next step is to to make a good looking GUI like
        Pokemon Ruby on GBA.''}
\end{itemize}
The other prompts were errands around the result, such as recording and hosting a
screencast.

\paragraph{Go-aheads.} A coding agent does not run a long plan to its end on its own. It
is built to stop at natural boundaries, such as the end of a planned phase, report what it
did, and wait for the human to confirm before it goes on, even when the plan is already
approved. The human then has to nudge it forward. Here the agent ran phases 0 to 6 of the
OS without a single stop, but stopped after every phase of the toolchain.
\ngoaheads{} of the \nprompts{} prompts are such nudges, of three words or fewer:
``go'', ``proceed'', ``continue'', ``Do T3.''. Each starts the next phase of a plan the
agent wrote; none adds information. The human also approved the agent's \napprovals{}
plans.

\paragraph{Decisions.} The human answered \nquestions{} multiple-choice questions from
the agent and dismissed \ndismissed{} (Table~\ref{tab:questions}). In \nrecommended{} of
them the human picked the option the agent had marked as recommended; only one answer was
free text. One recommended answer was overturned later. At the start, the human accepted
the agent's proposal to use the existing assembler and linker as tools, and the request
for a self-hosted toolchain reversed it. The human's own technical input was two prompts. One reported that typing in
\texttt{make run} showed nothing. The other questioned whether \texttt{objcopy} was needed
at all, which removed a planned phase (T6). One goal was dropped. When asked to run the
compiler inside the OS, the agent explained that the OS has no way to load native code,
and the human moved on to the game instead.

\begin{table}[t]
\caption{The human prompts of the session, verbatim, up to the request to play the game. The 7 later prompts are about recording and hosting screencasts and the graphical battle screen.}
\label{tab:prompts}
\begin{adjustbox}{max width=\linewidth}
\begin{tabular}{lp{11cm}}
\toprule
Time (CEST) & Prompt \\
\midrule
\href{\siteurl/t/42f0af3.html\#r-ab37688c}{Jul 17, 18:56} & I want to ask Claude to write a tiny OS with a tiny file system, with a tiny compiler and scripting language, all to be used in qemu/kvm or equivalent. the goal is that 100\% of lines of code come from Claude. Make a plan. \\
\href{\siteurl/t/24ac0a4.html\#r-5ebaacd1}{Jul 17, 22:10} & go \\
\href{\siteurl/t/cd6231d.html\#r-027250c3}{Jul 17, 22:12} & Now I want to use it, tell me how to boot and how to use \\
\href{\siteurl/t/cd6231d.html\#r-d1f83fd0}{Jul 17, 22:13} & when I type "make run", it starts, but when I type I don't see anything \\
\href{\siteurl/t/cd6231d.html\#r-67b0d2c6}{Jul 17, 23:02} & now make a plan to implement AS CC HOSTCC LD OBJCOPY and make such that we can build the system with generated tools only. The CC should compile the other tools and be self-hosted of course. \\
\href{\siteurl/t/72955af.html\#r-eaed13ce}{Jul 18, 08:01} & do T2/T4 \\
\href{\siteurl/t/f08a1b9.html\#r-928a41b6}{Jul 18, 08:37} & Do T3. \\
\href{\siteurl/t/1647d6f.html\#r-0e2d1d95}{Jul 18, 11:57} & Proceed with T5 \\
\href{\siteurl/t/1647d6f.html\#r-5c14eccb}{Jul 18, 16:40} & proceed \\
\href{\siteurl/t/1647d6f.html\#r-c6086a1f}{Jul 18, 16:49} & proceed \\
\href{\siteurl/t/0f82a8d.html\#r-0c3f5770}{Jul 18, 17:04} & About T6, I'm not sure it's required. There is no need to have OBJCOPY to build software. Is that spurious? Can we avoid it? \\
\href{\siteurl/t/0f82a8d.html\#r-4f1f7f69}{Jul 18, 17:06} & Do that. \\
\href{\siteurl/t/1d62d14.html\#r-f6ed2f35}{Jul 18, 17:14} & Do Phase T7. \\
\href{\siteurl/t/1d62d14.html\#r-927464f0}{Jul 18, 22:43} & continue \\
\href{\siteurl/t/acff0a5.html\#r-50d4a322}{Jul 18, 23:00} & Do Task T8. \\
\href{\siteurl/t/acff0a5.html\#r-4779aaf8}{Jul 19, 04:23} & go \\
\href{\siteurl/t/acff0a5.html\#r-9771e9a4}{Jul 21, 07:02} & proceed \\
\href{\siteurl/t/900c4cf.html\#r-d464bbfd}{Jul 21, 07:26} & o phase 9 \\
\href{\siteurl/t/4246ba7.html\#r-8565f973}{Jul 21, 18:58} & now install and test the self-hosted compiler inside tinyos. \\
\href{\siteurl/t/4246ba7.html\#r-229fe906}{Jul 22, 07:05} & implement in tinyos a Pokemon like game such that an AI agent later plays over serial port. \\
\href{\siteurl/t/377d751.html\#r-453af18e}{Jul 22, 11:37} & very good, start the game and play for 5 minutes, record the console screen cast of the game in a file. \\
\bottomrule
\end{tabular}
\end{adjustbox}
\end{table}

\begin{table}[t]
\caption{The multiple-choice questions the agent asked the human. Each question offered two to four options, one of them marked \emph{(Recommended)} by the agent; the human could also type free text. \emph{Dismissed}: the human rejected the question and changed the request.}
\label{tab:questions}
\begin{adjustbox}{max width=\linewidth}
\begin{tabular}{lp{7cm}p{5cm}}
\toprule
Time (CEST) & Question of the agent, verbatim & Answer of the human \\
\midrule
\href{\siteurl/t/42f0af3.html\#toolu_01RPMpMoUggkbKdae6xR4ACj}{Jul 17, 18:56} & Target boot platform for QEMU/KVM? & x86 BIOS, custom boot sector (Recommended) \\
\href{\siteurl/t/42f0af3.html\#toolu_01RPMpMoUggkbKdae6xR4ACj}{Jul 17, 18:56} & Implementation language for the OS/kernel/compiler source? & Assembly + freestanding C (Recommended) \\
\href{\siteurl/t/42f0af3.html\#toolu_01RPMpMoUggkbKdae6xR4ACj}{Jul 17, 18:56} & What counts as 'Claude writes 100\% of the code' regarding the toolchain? & Use existing assembler/linker/QEMU as tools (Recommended) \\
\href{\siteurl/t/42f0af3.html\#toolu_01RPMpMoUggkbKdae6xR4ACj}{Jul 17, 18:56} & How minimal should the OS itself be (scheduling/memory/drivers)? & Single-tasking, polling drivers (Recommended) \\
\href{\siteurl/t/cd6231d.html\#toolu_01VNQ4Z6pbFUGaFPacDCcLCX}{Jul 17, 23:06} & eval.c uses \~{}22 compound literals and codegen.c uses designated array initializers (C99 features). Implement these in the compiler, or refactor those two files to avoid them? & Refactor the 2 files (Recommended) \\
\href{\siteurl/t/cd6231d.html\#toolu_01VNQ4Z6pbFUGaFPacDCcLCX}{Jul 17, 23:06} & How rigorously should the compiler prove it's self-hosted? & Full 3-stage, byte-identical (Recommended) \\
\href{\siteurl/t/cd6231d.html\#toolu_01VNQ4Z6pbFUGaFPacDCcLCX}{Jul 17, 23:06} & Host-side tools (mkfs\_tinyfs) currently use libc (fopen/fread/strchr/exit). Once CC/LD replace gcc/ld, how should these be built? & Rewrite libc-free, raw Linux syscalls (Recommended) \\
\href{\siteurl/t/4246ba7.html\#toolu_016fLhCaCSLKdie5KoR5w6HN}{Jul 21, 18:59} & What does "install and test the self-hosted compiler inside tinyos" mean concretely? The self-hosted CC currently runs on the host (Linux) to build the OS — it doesn't run as a program inside the OS itself, and tinyos has no mechanism today to load/execute a native program from disk (the shell's `build`/`exec` only run the bytecode VM, not real machine code). & \emph{dismissed} \\
\bottomrule
\end{tabular}
\end{adjustbox}
\end{table}

\section*{Appendix: Phases of the Session}\label{app:phases}

Table~\ref{tab:phases} gives the speed of implementation phase by phase: the time, the
code written and the agent effort between two phase commits.

\begin{table}[t]
\centering
\caption{Phases of \sysname{}. \emph{Prompts} counts human messages, \emph{Active min} sums the gaps between transcript events shorter than 10 minutes, \emph{Output ktok} is thousands of tokens generated. Work between two phase commits is attributed to the later one; four minor commits are folded into the next phase.}
\label{tab:phases}
\begin{adjustbox}{max width=\linewidth}
\begin{tabular}{llrrrrrr}
\toprule
Commit (CEST) & Phase & Files & Lines added & Prompts & Active min & Output ktok & Tool calls \\
\midrule
\href{https://github.com/monperrus/TinyMonOS/commit/42f0af3}{Jul 17, 19:04} & \href{\siteurl/t/42f0af3.html}{0: boot sector, build pipeline} & 3 & 71 & 1 & 8 & 45 & 25 \\
\href{https://github.com/monperrus/TinyMonOS/commit/0b4c5c1}{Jul 17, 19:07} & \href{\siteurl/t/0b4c5c1.html}{1: two-stage loader, protected mode} & 10 & 350 & 0 & 3 & 26 & 20 \\
\href{https://github.com/monperrus/TinyMonOS/commit/1a67c58}{Jul 17, 19:10} & \href{\siteurl/t/1a67c58.html}{2: GDT/IDT, allocator, self-test} & 14 & 486 & 0 & 3 & 39 & 27 \\
\href{https://github.com/monperrus/TinyMonOS/commit/a3b80d1}{Jul 17, 19:16} & \href{\siteurl/t/a3b80d1.html}{3: ATA driver, tinyfs} & 13 & 620 & 0 & 6 & 69 & 24 \\
\href{https://github.com/monperrus/TinyMonOS/commit/594867b}{Jul 17, 19:20} & \href{\siteurl/t/594867b.html}{4: keyboard driver, shell} & 9 & 411 & 0 & 4 & 53 & 22 \\
\href{https://github.com/monperrus/TinyMonOS/commit/f4702ba}{Jul 17, 19:30} & \href{\siteurl/t/f4702ba.html}{5: scripting language} & 16 & 1126 & 0 & 10 & 135 & 49 \\
\href{https://github.com/monperrus/TinyMonOS/commit/3c37ed4}{Jul 17, 19:38} & \href{\siteurl/t/3c37ed4.html}{6: compiler, bytecode VM} & 13 & 1248 & 0 & 7 & 109 & 30 \\
\href{https://github.com/monperrus/TinyMonOS/commit/24ac0a4}{Jul 17, 22:12} & \href{\siteurl/t/24ac0a4.html}{7: README, docs, final check} & 2 & 181 & 1 & 2 & 15 & 12 \\
\midrule
\href{https://github.com/monperrus/TinyMonOS/commit/cd6231d}{Jul 17, 23:40} & \href{\siteurl/t/cd6231d.html}{T1: assembler} & 11 & 2053 & 3 & 46 & 451 & 126 \\
\href{https://github.com/monperrus/TinyMonOS/commit/72955af}{Jul 18, 08:36} & \href{\siteurl/t/72955af.html}{T2/T4: C compiler} & 21 & 2999 & 1 & 35 & 474 & 143 \\
\href{https://github.com/monperrus/TinyMonOS/commit/f08a1b9}{Jul 18, 08:49} & \href{\siteurl/t/f08a1b9.html}{T3: adapt kernel to tinyC} & 18 & 357 & 1 & 13 & 100 & 75 \\
\href{https://github.com/monperrus/TinyMonOS/commit/1647d6f}{Jul 18, 16:54} & \href{\siteurl/t/1647d6f.html}{T5: linker} & 15 & 1101 & 3 & 57 & 427 & 263 \\
\href{https://github.com/monperrus/TinyMonOS/commit/0f82a8d}{Jul 18, 17:11} & \href{\siteurl/t/0f82a8d.html}{T6: drop \texttt{objcopy}} & 4 & 66 & 2 & 17 & 44 & 34 \\
\href{https://github.com/monperrus/TinyMonOS/commit/1d62d14}{Jul 18, 22:59} & \href{\siteurl/t/1d62d14.html}{T7: switch the build over} & 20 & 1022 & 2 & 110 & 1\,065 & 458 \\
\href{https://github.com/monperrus/TinyMonOS/commit/acff0a5}{Jul 21, 07:17} & \href{\siteurl/t/acff0a5.html}{T8: self-hosting bootstrap} & 30 & 1131 & 3 & 86 & 724 & 470 \\
\href{https://github.com/monperrus/TinyMonOS/commit/900c4cf}{Jul 21, 07:28} & \href{\siteurl/t/900c4cf.html}{T9: docs, cleanup} & 1 & 23 & 1 & 12 & 13 & 15 \\
\midrule
\href{https://github.com/monperrus/TinyMonOS/commit/4246ba7}{Jul 22, 07:20} & \href{\siteurl/t/4246ba7.html}{serial-playable battle game} & 15 & 664 & 2 & 17 & 114 & 71 \\
\href{https://github.com/monperrus/TinyMonOS/commit/377d751}{Jul 22, 16:56} & \href{\siteurl/t/377d751.html}{VGA mode 13h, bitmap font} & 6 & 429 & 6 & 69 & 326 & 171 \\
\href{https://github.com/monperrus/TinyMonOS/commit/a4a64cb}{Jul 22, 17:21} & \href{\siteurl/t/a4a64cb.html}{graphical battle screen} & 10 & 520 & 0 & 25 & 180 & 107 \\
\midrule
after last commit & \href{\siteurl/t/after.html}{screencasts, hosting} & ---& --- & 2 & 22 & 162 & 72 \\
\bottomrule
\end{tabular}
\end{adjustbox}
\end{table}

\end{document}